\documentstyle[12pt]{article}

\catcode`\@=11
\long\def\@makefntext#1{
\protect\noindent \hbox to 3.2pt {\hskip-.9pt
$^{{\ninerm\@thefnmark}}$\hfil}#1\hfill}		

\def\@makefnmark{\hbox to 0pt{$^{\@thefnmark}$\hss}}  

\def\ps@myheadings{\let\@mkboth\@gobbletwo
\def\@oddhead{\hbox{}
\rightmark\hfil\ninerm\thepage}
\def\@oddfoot{}\def\@evenhead{\ninerm\thepage\hfil
\leftmark\hbox{}}\def\@evenfoot{}
\def\sectionmark##1{}\def\subsectionmark##1{}}

\renewcommand{\thefootnote}{\fnsymbol{footnote}}

\newcounter{sectionc}\newcounter{subsectionc}\newcounter{subsubsectionc}
\renewcommand{\section}[1] {\vspace*{0.6cm}\addtocounter{sectionc}{1}
\setcounter{subsectionc}{0}\setcounter{subsubsectionc}{0}\noindent
	{\normalsize\bf\thesectionc. #1}\par\vspace*{0.4cm}}
\renewcommand{\subsection}[1] {\vspace*{0.6cm}\addtocounter{subsectionc}{1}
	\setcounter{subsubsectionc}{0}\noindent
	{\normalsize\it\thesectionc.\thesubsectionc. #1}\par\vspace*{0.4cm}}
\renewcommand{\subsubsection}[1]
{\vspace*{0.6cm}\addtocounter{subsubsectionc}{1}
	\noindent {\normalsize\rm\thesectionc.\thesubsectionc.\thesubsubsectionc.
	#1}\par\vspace*{0.4cm}}

\newcounter{appendixc}
\newcounter{subappendixc}[appendixc]
\newcounter{subsubappendixc}[subappendixc]

\renewcommand{\appendix}[1] {\vspace*{0.6cm}
        \refstepcounter{appendixc}
        \setcounter{figure}{0}
        \setcounter{table}{0}
        \setcounter{equation}{0}
        \renewcommand{\thefigure}{\Alph{appendixc}.\arabic{figure}}
        \renewcommand{\thetable}{\Alph{appendixc}.\arabic{table}}
        \renewcommand{\theappendixc}{\Alph{appendixc}}
        \renewcommand{\theequation}{\Alph{appendixc}.\arabic{equation}}
        \noindent{\bf Appendix \theappendixc #1}\par\vspace*{0.4cm}}

\def\abstracts#1{{

\centering{\begin{minipage}{12.2truecm}\footnotesize\baselineskip=12pt\noindent
	\centerline{\footnotesize ABSTRACT}\vspace*{0.3cm}
	\parindent=0pt #1
	\end{minipage}}\par}}

\renewenvironment{thebibliography}[1]
	{\begin{list}{\arabic{enumi}.}
	{\usecounter{enumi}\setlength{\parsep}{0pt}
\setlength{\leftmargin 1.25cm}{\rightmargin 0pt}
	 \setlength{\itemsep}{0pt} \settowidth
	{\labelwidth}{#1.}\sloppy}}{\end{list}}

\newcounter{itemlistc}
\newcounter{romanlistc}
\newcounter{alphlistc}
\newcounter{arabiclistc}

\newcommand{\fcaption}[1]{
        \refstepcounter{figure}
        \setbox\@tempboxa = \hbox{\footnotesize Fig.~\thefigure. #1}
        \ifdim \wd\@tempboxa > 6in
           {\begin{center}
        \parbox{6in}{\footnotesize\baselineskip=12pt Fig.~\thefigure. #1}
            \end{center}}
        \else
             {\begin{center}
             {\footnotesize Fig.~\thefigure. #1}
              \end{center}}
        \fi}

\newcommand{\tcaption}[1]{
        \refstepcounter{table}
        \setbox\@tempboxa = \hbox{\footnotesize Table~\thetable. #1}
        \ifdim \wd\@tempboxa > 6in
           {\begin{center}
        \parbox{6in}{\footnotesize\baselineskip=12pt Table~\thetable. #1}
            \end{center}}
        \else
             {\begin{center}
             {\footnotesize Table~\thetable. #1}
              \end{center}}
        \fi}

\def\@citex[#1]#2{\if@filesw\immediate\write\@auxout
	{\string\citation{#2}}\fi
\def\@citea{}\@cite{\@for\@citeb:=#2\do
	{\@citea\def\@citea{,}\@ifundefined
	{b@\@citeb}{{\bf ?}\@warning
	{Citation `\@citeb' on page \thepage \space undefined}}
	{\csname b@\@citeb\endcsname}}}{#1}}

\newif\if@cghi
\def\cite{\@cghitrue\@ifnextchar [{\@tempswatrue
	\@citex}{\@tempswafalse\@citex[]}}
\def\citelow{\@cghifalse\@ifnextchar [{\@tempswatrue
	\@citex}{\@tempswafalse\@citex[]}}
\def\@cite#1#2{{$\null^{#1}$\if@tempswa\typeout
	{IJCGA warning: optional citation argument
	ignored: `#2'} \fi}}

 1
 1
 1

\font\ninerm=cmr9

\begin{document}

\newcommand{\st}{\scriptstyle}
\newcommand{\sst}{\scriptscriptstyle}
\newcommand{\mco}{\multicolumn}
\newcommand{\epp}{\epsilon^{\prime}}
\newcommand{\vep}{\varepsilon}
\newcommand{\ra}{\rightarrow}
\newcommand{\ppg}{\pi^+\pi^-\gamma}
\newcommand{\vp}{{\bf p}}
\newcommand{\ko}{K^0}
\newcommand{\kb}{\bar{K^0}}
\newcommand{\al}{\alpha}
\newcommand{\ab}{\bar{\alpha}}
\def\be{\begin{equation}}
\def\ee{\end{equation}}
\def\bea{\begin{eqnarray}}
\def\eea{\end{eqnarray}}
\def\CPbar{\hbox{{\rm CP}\hskip-1.80em{/}}}

\centerline{\normalsize\bf POSSIBLE HIGGS BOSON EFFECTS ON THE RUNNING
OF THIRD}
\centerline{\normalsize\bf AND FOURTH GENERATION QUARK MASSES AND MIXINGS}
\baselineskip=15pt

\centerline{\footnotesize LESLEY L. SMITH\footnote{Conference speaker}}
\baselineskip=13pt
\centerline{\footnotesize\it Department of Physics and Astronomy,
University of Kansas, Lawrence, KS 66045, U.S.A.}
\baselineskip=13pt
\centerline{\footnotesize E-mail: smith@kuphsx.phsx.ukans.edu}
\vspace*{0.3cm}
\centerline{\footnotesize PANKAJ JAIN}
\baselineskip=13pt
\centerline{\footnotesize\it Department of Physics and Astronomy, University of
Oklahoma, Norman, OK, 73019, U.S.A.}
\vspace*{0.3cm}
\centerline{\footnotesize DOUGLAS W. MCKAY}
\baselineskip=13pt
\centerline{\footnotesize\it Department of Physics and Astronomy, University of
Kansas, Lawrence, KS, 66045, U.S.A.}
\vspace*{0.9cm}
\abstracts{The Schwinger-Dyson equation for the quark self-energy is solved
for the case of the third and fourth quark generations.  The exchanges of
standard model gluons and Higgs bosons are taken into account.  It is found
that Higgs boson exchange dominates the quark self-energy in the ultraviolet
region for sufficiently large input quark masses causing the running quark
propagator mass to increase with energy-scale.  No running of the quark mixing
angles is found for input quark masses up to and including 500 GeV.}

\normalsize\baselineskip=15pt
\setcounter{footnote}{0}
\renewcommand{\thefootnote}{\alph{footnote}}

\indent
\par
Unlike the renormalization group equation,
the Schwinger-Dyson equation
 (SDE) analysis presented here
enables us in principle to calculate the dynamical quark mass functions
for very light quarks at low energies
where quantum chromodynamics (QCD) is in the non-perturbative region.
This is of interest because
experimentally accessible quantities are presently found at low
energy-scales.  In the presence of heavy quarks, such as bottom and top,
it allows the calculation of running mass matrix
at momentum scale of the order or less than the quark masses, at which
scale the RG equation results may not be reliable.
Moreover,
 this nonperturbative SDE approach is also interesting in that it can
describe the running mass function of an ultra-heavy quark, with a large
 Yukawa coupling that cannot be treated reliably by perturbation
theory.  This would be relevant for a fourth
generation quark.
\par
In this study we
assume that chiral symmetry is broken and
solve the quark propagator SDE to find running quark mass functions
and running quark mixing angles.  In addition to the usual gluon
contributions, we include Higgs boson exchange effects
in the quark self-energy.  The
contributions to the quark self-energy due to Higgs bosons are being studied
here for the first time in the context of the quark propagator SDE.  We also
study for the first time in a Higgs-boson-plus-gluon model the
multigenerational
cases of quarks which enable us to analyze quark mixing angles.
\indent
\par The general form of the quark propagator SDE that we use is
\begin{equation}
S^{-1}(q)=S^{-1}_{0}(q)-\int \frac{d^{4}k}{(2\pi)^{4}} g_s\gamma_{\mu}
S(k)g_{s}\Lambda_{\nu}G^{\mu\nu}-\int \frac{d^{4}k}{(2\pi)^{4}} g_{H}
S(k)g_{Y}\Lambda_{H}P_{H}, \label{SDE}
\end{equation}
where $g_s\gamma_{\mu}$ and $g_{s}\Lambda_{\nu}$ are quark-gluon vertices,
$S(k)$ is the quark propagator, $G^{\mu\nu}$ is the gluon propagator,
$g_{Y}\Lambda_{H}$ is the quark-Higgs boson vertex factor, and $P_{H}$
is the Higgs boson propagator.
\par
The ladder
approximation quark propagator in the Landau gauge in Minkowski space that
we use is $iS(k)=({\not}{k}+M)(k^{2}-M^{2})^{-1}$, where $M$ is a
$2\times2$ quark mass function matrix, assumed to be symmetric for this
study.
The gluon propagator we use is
\begin{equation}
g^{2}_{s}G^{\mu\nu}(k) \equiv i\left(-g^{\mu\nu}+
\frac{k^{\mu}k^{\nu}}{k^{2}} \right)
G(k), \hspace{.5cm}
{\rm with} \hspace{.25cm}
G(k)=\frac{g^{2}_{s}(k)}{k^{2}}
\end{equation}
and the leading-log (one-loop) coupling
\begin{equation}
g^{2}_{s}(k)=
\frac{4\pi^{2}d}{ln(x_{0}-\frac{k^{2}}{\Lambda^{2}_{QCD}})}
\label{gQCD}
\end{equation}
where $d=12/(33-2n_{f})$, $n_{f}$ is the number of flavors.
This leading-log QCD
coupling is renormalization-scheme and gauge independent, as is its second
order, two-loop, extension \cite{twoloop}.  $x_{0}$ is a parameter
that functions as a smooth infrared cutoff.  Eq. (\ref{gQCD}) does not
accurately model the gluon potential in the infrared
region, but it does enable
us to assess the relative magnitude and shape effects
of the Higgs boson and gluon exchanges in the quark self-energy \cite{note3}.
\par For the Yukawa sector we consider a $2\times2$ symmetric
Yukawa coupling matrix, which can be diagonalized by a $2\times2$ unitary
matrix, to give diagonal elements $g_{\pm}$.  The one-loop Renormalization
Group Equation for the diagonalized Yukawa couplings, $g_{\pm}$, without
weak interaction effects, is
\begin{equation}
\frac{dg_{\pm}}{dt}=\frac{1}{4\pi^{2}}
\left( \frac{9}{8}g^{3}_{\pm}(t)-2g^{2}_{QCD}(t)g_{\pm}(t) \right)
\end{equation}
with $t=ln(x_{0}+q^{2}/\Lambda^{2}_{QCD})/2$, and $\Lambda_{QCD}=$ 0.18 GeV.
This has solution
\begin{equation}
g^{2}_{\pm}=\frac{1}{C_{\pm}t^{2d}-\frac{9}{16\pi^{2}}\frac{t}{1-2d}}
\label{gpm}
\end{equation}
where the $C_{\pm}$ must be determined by boundary conditions.
We apply the boundary
conditions
$m^{2}_{\pm}(1\; {\rm GeV})=g^{2}_{\pm}(1\;{\rm GeV}){\rm v}^{2}/2$
for small ($<150$ GeV) running quark masses, and
$m^{2}_{\pm}(m_{\pm})=g^{2}_{\pm}(m_{\pm}){\rm v}^{2}/2$ for large
running quark
masses, where v $\simeq$ 246 GeV.
\par We solve the resulting SDE
by converting it into a second order differential
equation and applying a fourth-order Runge-Kutta subroutine \cite{Smith95}.
We input $M_{bottom}(0)=$4.89 GeV, $M_{top}(0)=$179 GeV, and
for the degenerate fourth generation quarks
$M_{bottom'}(0)=M_{top'}(0)=$506 GeV \cite{Munczek,PDB94}.
For the third-fourth two-quark-generation case we input the following
$q^{2}/\Lambda^{2}_{QCD}=0$ values of the mixing angles
for the up-sector and down-sector:
 $\theta_{up}=0.5$ radians, and
$\theta_{down}=\theta_{up}-\theta_{cabibbo}$ with
$\theta_{cabibbo}=0.22$ radians \cite{PDB94}.
We are primarily interested in the question
of the running of the mixing angles.

\par  The QCD plus Higgs boson case of the bottom quark mass function does not
deviate significantly from the pure QCD running mass function.
The QCD
plus Higgs boson case of the top quark mass function shows some deviation
from pure QCD.
The Higgs boson term has an opposite sign relative to the gluon term, so it
drives the top mass function up, in contrast to the usual QCD-only decreasing
 mass function.
  The Higgs boson has a
substantial effect on the fourth generation quark self-energy.
In fact, it causes the running fourth generation quark masses to
increase in the asymptotic
region.  The results are summarized in the table, where $A$ and $B$ are the
initial and final integration points.

\begin{table}\begin{center}
\tcaption{Two-quark-generation mass results, for
$A=5 \times 10^{-5}$, $B=9.77 \times 10^{8}$.}

\begin{tabular}{|c||c|c|c|} \hline
quark & $M_{q}(A)$ & $M_{q}(B)$ & $\frac{M_{QCD+H}}{M_{QCD}}$ \\ \hline\hline
bottom & 27.2$\Lambda_{QCD}$ & 15.3$\Lambda_{QCD}$ & 1 \\ \hline
top & 993$\Lambda_{QCD}$ & 840$\Lambda_{QCD}$ & 1.01 \\ \hline
bottom$'$=top$'$ & 2810$\Lambda_{QCD}$ & 3150$\Lambda_{QCD}$ &
$\neq$1, varies \\ \hline\hline
\end{tabular}
\end{center}
\end{table}

\par  By studying the mixing angle results, we determine that the running
of the mixing angles due to any effects in our model is less
than one part in $10^{9}$, for any quark mass input value up to and
including 500 GeV.
This is expected for pure QCD and can be proven analytically for QCD
\cite{Smith}.
  We have been unable to prove
analytically that the cabibbo angle is constant when the Yukawa interaction is
included in the SDE analysis, but we have established numerically
that there can be no observable effect even for the heavy hypothetical fourth
generation.


\end{document}